%% file: ODI.tex
\documentclass[aps,prl,reprint,superscriptaddress,twocolumn,hidelinks]{revtex4-2}

\usepackage{graphicx}

\usepackage{rotating}

\usepackage{dcolumn}

\usepackage{amssymb}

\usepackage{amsmath}

\usepackage{bm}

\usepackage{listings}

\usepackage{fancyhdr}

\usepackage{wrapfig}

\usepackage{subfig}

\usepackage{xcolor}

\usepackage{hyperref}

\hypersetup{
    colorlinks,
    linkcolor={red!50!black},
    citecolor={blue!50!black},
    urlcolor={blue!80!black}
}

\include{mydefs}

\allowdisplaybreaks

\begin{document}

\title{Large gauge transformations, local coordinates and cosmological observables}

\author{Ermis Mitsou}
\email[]{ermitsou@physik.uzh.ch}
\affiliation{Center for Theoretical Astrophysics and Cosmology, Institute for Computational Science, University of Zurich, CH--8057 Zurich, Switzerland}

\author{Jaiyul Yoo}
\email[]{jyoo@physik.uzh.ch}
\affiliation{Center for Theoretical Astrophysics and Cosmology, Institute for Computational Science, University of Zurich, CH--8057 Zurich, Switzerland}
\affiliation{Physics Institute, University of Zurich, Winterthurerstrasse 190, CH--8057, Zurich, Switzerland}

\date{\today}

\begin{abstract}
In recent years new types of coordinate transformations have appeared in cosmology on top of the standard gauge transformations, such as the dilatations and special conformal transformations, or the ones leading to (conformal) Fermi coordinates. Some of these can remove effects that are invariant under the standard gauge transformations and also affect asymptotic boundary conditions, thus introducing a non-trivial ambiguity in our cosmological modeling. In this short note we point out that this ambiguity is irrelevant for the quantities we use to compare our model with observations -- the cosmological observable relations -- as they are {\it invariant} under {\it all} of these transformations. Importantly, this invariance holds only if one takes into account {\it all} the relativistic contributions to an observable, which is not the case in the literature in general. We finally also show that the practically-relevant property of conformal Fermi coordinates (a FLRW metric up to second order in distance) can be achieved through a globally-defined standard gauge transformation. 
\end{abstract}

\maketitle

{\bf Introduction} -- Many elements of our current cosmological modeling lack an explicitly coordinate-independent description (e.g. initial conditions, field statistics or tracer-matter bias models). This is because they are defined within cosmological perturbation theory, thus inheriting a dependence on the background solution and an implicit restriction to a subclass of global coordinate systems -- those in which the metric is globally described by small deviations from the background. Within this restricted set, general covariance manifests itself as the usual ``small'' gauge transformations (SGTs), i.e. the coordinate transformations that relate globally ``small'' metric fluctuations (or, equivalently, relating different background--fluctuation decompositions). In particular, physical information must be SGT-invariant. 

In recent years, however, more general types of coordinate transformations have been shown to be useful, i.e. spoiling either the global smallness of the metric fluctuations, or the fact that they can even be described globally. These are diffeomorphisms that do not reduce to the identity at infinity, known as ``large'' gauge transformations (LGT), or transformations to coordinate systems with a finite domain of definition, usually simply referred to as ``local'' coordinates (see below for examples). In particular, they can alter SGT-invariant quantities. As a result, they force upon us the challenging task of identifying physical quantities and relations, i.e. those that are {\it fully} coordinate-independent, within a model that privileges a coordinate class in its very definition. 

Perhaps the best illustration of this issue arises in galaxy bias, where one models the relation between the matter field and the galaxies that act as observable tracers thereof. This relation is governed by very localized processes compared to cosmological scales, so the global coordinate systems used in cosmology (e.g. conformal-Newtonian) never coincide with the physical distances involved. They therefore always carry long-wavelength coordinate artifacts, which precludes a proper modeling. The most notorious example is the non-local contribution to galaxy bias induced by single-field primordial non-Gaussianity, which is indistinguishable from a LGT within a finite volume (see \cite{dePutter:2015vga} and references therein). To avoid this issue one must eliminate these artifacts, i.e. escape the subclass of global coordinate systems through some LGT or local coordinate system. The simplest choice is the Fermi normal coordinates \cite{Baldauf:2011bh}, which satisfy $g_{\mu\nu} = \et_{\mu\nu} + \Ord(\vec{x}^2)$ over the galaxy region and thus coincide with the physical local distances up to small controllable corrections.

If we had an explicitly fully coordinate/background-independent bias model it would not matter which coordinate system class is chosen to do computations and one could describe these local physics directly in global coordinates. In the absence of such a technology, however, the only option is to complement the existing model by an ad hoc prescription to use different coordinate classes for global and local processes. The problem of such constructions is that they introduce an artificial notion of ``right'' and ``wrong'' coordinate system and this can lead to confusion. Since SGT-invariance is not preserved under general coordinate transformations, there is no criterion for knowing whether one has the ``correct'' result, or whether some LGT or local coordinate is actually needed. 

Note that background-dependence, and the resulting distinction of ``gauge classes'' at the level of the model, is also very common in other fields such as high-energy physics. In that context the analogues of LGTs have a non-trivial physical effect, which will not be the case in cosmology, as we will see. The reason for this is a different definition of what constitutes an ``observable'' in each case. Note, finally, that in gauge theory the terms ``SGT'' and ``LGT'' are mostly employed for something else: the gauge transformations that are homotopic and non-homotopic to the identity, respectively.

{\bf Goal} -- The aim of this short note is to stress that there is one part of our cosmological modeling that admits a fully background/coordinate-independent description and therefore does not require any extra prescriptions in its definition: the cosmological observable relations. We first discuss the most important types of LGTs and local coordinate systems in the literature in order to clarify their impact. We then move on to explain why cosmological observable relations are completely invariant under {\it any} coordinate transformation, we point out the conceptual differences with the case of LGTs in high-energy physics contexts and we conclude with a note of caution for explicit computations. Finally, we also show that one can achieve the advantages of local coordinates with an SGT, i.e. without spoiling the validity of global cosmological perturbation theory.

{\bf Residual gauge freedom ``at infinity''} -- The utility of LGTs in cosmology was highlighted by the discovery of conserved infrared modes and consistency relations for cosmological correlation functions \cite{Maldacena:2002vr,Weinberg:2003sw,Creminelli:2004yq,Cheung:2007sv,Creminelli:2012ed,Hinterbichler:2013dpa,Creminelli:2013mca}. The LGTs of interest in these cases arise as symmetries of a fully SGT-fixed metric, i.e. diffeomorphisms that preserve that particular metric form, and thus constitute a residual freedom on top of SGTs. The prototypical example is the constant dilatation $x^i \to e^{\la} x^i$ in the $\ze$-gauge $g_{ij}(t,\vec{x}) = a^2(t) \, e^{2\ze(t,\vec{x})} \de_{ij}$, which preserves the form, but shifts the potential $\ze(\vec{x}) \to \ze(e^{\la} \vec{x}) + \la$. This transformation is ambiguous at the level of scalar-vector-tensor components \cite{Matarrese:2020why}. Indeed, the generating field is a pure-scalar $\xi^i = \la x^i \equiv \la \pa_i \( \vec{x}^2/2 \)$, so the SGT implementation would generate a $\sim \pa_i \pa_j$ metric component, but this leads to the same result as shifting the $\sim \de_{ij}$ one since $\pa_i \pa_j (\vec{x}^2/2) \equiv \de_{ij}$. Therefore, one can consider any weighted average of these two options. This ambiguity occurs precisely because $\xi^i$ is not bounded in space: the metric variation has a singular Fourier transform at $\vec{k} = 0$, i.e. $\propto \de_{ij} \de^{(3)}(\vec{k})$, so its scalar-vector-tensor decomposition, which involves negative $k$ powers, is not unique. In contrast, SGTs do not suffer from this issue, because their generating vectors must tend to zero at infinity.

In the case of purely scalar perturbations the symmetry also extends to special conformal transformations, which shift the potential $\ze$ by $\sim x^i$ to lowest order \cite{Creminelli:2012ed,Hinterbichler:2012nm}, while the tensor sector admits an infinite set of such transformations \cite{Hinterbichler:2012nm}. Forming a discrete set with constant parameters, these LGTs are therefore global symmetries of the SGT-fixed metric form \footnote{A subtlety: The residual LGTs turn solutions with trivial boundary conditions into solutions with non-vanishing boundary conditions. The latter will not necessarily be physical, in the sense that they are not limits of solutions with trivial boundary conditions, but this can be corrected by combining with appropriate time and space translations \cite{Weinberg:2003sw,Creminelli:2012ed,Hinterbichler:2012nm}. These properly complemented LGTs then map physical solutions to physical solutions.}. Unlike translations and rotations, however, they transform the metric components non-linearly, i.e. as a gauge transformation would, and thus allow one to eliminate specific profiles (e.g. the constant mode of $\ze$). Moreover, since these LGTs appear as a residual freedom to SGTs, they can alter part of SGT-invariant quantities (e.g. the constant in $\ze$).

{\bf Local coordinate systems} -- Another type of non-trivial coordinate transformations are those that simplify the metric locally around a given geodesic world-line, so that the resulting coordinates are closely related to physical distances locally. The case where this relation is an identity are the Fermi normal coordinates (FC), which are built through a 3D exponential map and a choice of a rest-frame triad along the world-line, but whose relevant property here is $g_{\mu\nu} = \et_{\mu\nu} + \Ord(\vec{x}^2)$ around the world-line. The FC were then generalized to the `conformal Fermi coordinates' (CFC) \cite{Pajer:2013ana,Dai:2015rda,Dai:2015jaa}, satisfying instead 
\beq \label{eq:CFC}
g_{\mu\nu}^{\rm CFC} = a^2(t) \[ \et_{\mu\nu} + \Ord(\vec{x}^2) \] \, , 
\eeq
i.e. the metric is now the trivial cosmological solution (FLRW) up to second order in distance, instead of flat space-time. The CFC is more convenient for cosmological applications, because the $\Ord(\vec{x}^{n\geq 2})$ corrections solely depend on fluctuations, not on background functions as in the FC case, and can thus be controlled at all times. 

The transformation to (C)FC is constructed order by order in $x^i$, which obscures the behaviour at spatial infinity. Depending on the solution, it could be a LGT and/or its scalar-vector-tensor decomposition could be ill-defined, so that its effect on SGT-invariant quantities is unclear. In fact, in the generic case the (C)FC is local, i.e. it can only cover a finite patch of space-time \footnote{In the FC construction this occurs when two distinct space-like geodesics starting at the same point on the world-line cross each other again at some other point (two Fermi coordinates for the same point), or when there exist points that cannot be connected by any geodesic to the world-line. The CFC case inherits similar limitations.}, so its behavior at spatial infinity is not even defined. 

In practice, however, the (C)FC line-element is always used up to finite order in $x^i$, i.e. within a finite volume, so it is effectively just another type of LGT. Moreover, one usually only transforms the long-wavelength part of the metric to the form \eqref{eq:CFC}, meaning that the short wavelengths are still described by a global coordinate system. The purpose is similar to the one of the residual LGTs, i.e. to eliminate long-wavelength contributions to the metric by simplifying the first few coefficients in a spatial expansion. The CFC idea actually amounts to taking that particular property to the extreme (see \eqref{eq:CFC}), at the expense of altering the gauge, i.e. transforming to CFC is not a residual freedom on top of SGTs.

{\bf Cosmological observable relations} -- An important aspect of the works studying the effects of LGTs and (C)FC is that they focus on quantities in real or Fourier space, such as correlation functions and (multi-)spectra. In this case, even if one works with SGT-invariant fields, these quantities are still coordinate-dependent through the very use of functions of $x^i$ or the dual $k^i$, and can thus vary under the residual LGTs or when transforming to (C)FC. However, what we actually measure are cosmological observable relations, such as the average luminosity distance as a function of redshift $\bra D_L \ket(z)$, or the CMB temperature as a function of incoming photon direction $T_{\rm CMB}(\hat{n})$. These make no mention of space-time coordinates whatsoever. They only require an orthonormal basis of the tangent space at the observer point (a ``tetrad'') \cite{Schmidt:2012ne,Mitsou:2019nhj}, independently of any local coordinate system. They therefore admit an explicitly coordinate/background-independent definition through the tetrad formulation of differential geometry \cite{Mitsou:2019nhj}. 

Let us summarize this construction. First, one picks an integral line of the time-like tetrad vector field and interprets it as the observer's world-line. Therefore, along that line, the time-like tetrad vector is the 4-velocity of the observer, while the space-like tetrad vectors are interpreted as the spatial frame that this observer uses to do measurements. Since this information is expressed covariantly, i.e. through vector fields, no local coordinate system needs to be invoked to induce an orthonormal frame at the observer (as FC would for instance). Away from the world-line the tetrad field is interpreted as the 4-velocities and (fictitious) spatial frames of sources. Changing this observer/source data can then be achieved through the local Lorentz symmetry of the tetrad formulation. Next, one considers all the past light-like geodesics $\ga_{\ta,\hat{n}}$ emanating from the observer world-line, where $\ta$ is the proper time of observation and $\hat{n}$ is the observed angle in the sky. Each geodesic path is then parametrized by the photon's redshift $\ga_{\ta,\hat{n}}^{\mu}(z)$. The set $\{ \ta, z, \hat{n} \}$ is a particular case of  ``observational coordinates''  \cite{ObsCoord} and is defined with respect to the observer/source frames, so it is coordinate-independent. The image of the map $\ga^{\mu}(\ta,\hat{n},z)$ is the observable universe of the observer, by definition. Pulling back the metric along $\ga$ we obtain optical-type observable relations (e.g. $D_A(\ta, z, \hat{n})$), while source-related observables are found in two steps: first we project fields and phase space distributions in the tetrad basis to obtain diffeomorphism scalars \cite{Mitsou:2019nhj,Yoo:2017svj} and then pull back along $\ga$. In both cases we are left with explicitly coordinate/background-independent relations of the form $\Ord(\ta,z,\hat{n})$, which is therefore an example of so-called ``relational observables'' (for a discussion of the concept and references, see \cite{Goeller:2022rsx}).

Finally, note that the (C)FC idea can also be implemented as a construction of observer-field relational observables \cite{Urakawa:2010it,Tanaka:2011aj}, since in that case too one shoots geodesics from an observer world-line. One would simply need to keep the corresponding map $\ga^{\mu}(\ta,\vec{x}_{\rm FC})$ in generic target coordinates, e.g. to simplify the resolution of equations of motion for instance, so that the $\{ \ta, x_{\rm FC}^i \}$ data are interpreted as a coordinate-independent label of events by the observer. The important difference is that these are space-like geodesics, instead of light-like, so the corresponding observables are not {\it cosmological} observables. In the CFC-related literature they are usually referred to as ``local observables'' and equation \eqref{eq:CFC} are the coordinates an ``observer'' at $x^i = 0$ would use to ``measure'' them. To avoid the potential confusion that this similar terminology can bring in the present context, we repeat that these are not the ``cosmological observables'' that the actual ``observer'' is really ``measuring'' with cosmological surveys. Instead, the local observables in the CFC context are physical quantities that a spatially-extended set of fictitious observers around $x^i = 0$ could measure in principle (e.g. when studying local processes such as galaxy formation), since equation \eqref{eq:CFC} is simplifying the metric in space-like directions. Instead, for cosmological observables, the observer at $x^i$ can only access the information on their past light-cone, i.e. along light-like directions instead of space-like. In that case a more adapted coordinate system is the analogue of equation \eqref{eq:CFC}, but in light-cone adapted coordinates, as explicitly constructed in \cite{Mitsou:2020czr}. Although the CFC literature clearly mentions the need for an extra computational step to construct the cosmological observables out of CFC data, the use of similar terminology could lead to confusion, and in particular to the erroneous conclusion that CFC is related to, or even needed for, cosmological observables.  

{\bf Distinction from LGTs in high-energy physics} -- Let us now discuss the difference with the use and interpretation of LGTs in the cases of typical interest in high-energy physics. In that context one considers some ``background'' solution and fluctuations with trivial boundary conditions at its conformal (asymptotic) boundary. This way the bulk system can be unambiguously described through quantities that are defined asymptotically and with respect to that background (e.g. radiation, charges, amplitudes, etc.). This is what observers at ``infinity'' (the boundary) would measure, which therefore endows these quantities with a physical interpretation -- they are the ``observables'' of the system. The set of symmetries is then split into those which preserve the physical quantities, i.e. the analogues of SGTs here, and the rest, which are therefore the analogues of LGTs \footnote{An example in the case of gravity, which attracted much attention recently, are the LGTs affecting the asymptotic behaviour of the metric at null infinity in a Minkowski background and forming the Bondi–Metzner–Sachs group (see \cite{Strominger:2017zoo} and references therein). These can also be generalized to the case of spatially-flat decelerating FLRW background \cite{Enriquez-Rojo:2020miw} (accelerating FLRW has a conformal boundary at time-like infinity, instead of null).}. The latter are thus also endowed with a physical interpretation, because altering observable quantities implies that they relate physically-distinct solutions. 

The situation is different in cosmology, where one is interested in constraining model parameters with actual observations, i.e. taking place from within the system, instead of a boundary. This qualitative difference is also forced upon us by a standard ingredient of cosmological modelling: the statistical homogeneity and isotropy of the stochastic fluctuations. Indeed, although the use of the Fourier transform and scalar-vector-tensor decomposition implicitly assumes a trivial behaviour at spatial infinity, that boundary loses its privileged status once we impose translational and rotational invariance on the statistics. The observer is therefore unavoidably ``in the middle'' of the system, instead of ``at infinity'', and interacts with the system, instead of receiving information from afar. Thus, the ``observables'' in this case (cosmological observable relations) are not defined asymptotically with respect to some background, but through a relation in the bulk, one between fields in a finite region of space-time and a framed observer world-line therein. Since the latter is related to the former covariantly (e.g. non-precessing free-fall), both entities transform together under {\it any} diffeomorphism and preserve the observable relations. Consequently, even if two solutions related by a LGT are interpreted as physically distinct by a high-energy theorist, they are still observationally indistinguishable to a cosmologist. Put differently, it does not matter how an LGT would affect fields at infinity, because the corresponding {\it cosmological} observables would be invariant.

{\bf A note of caution} -- Given the invariance of observable relations under all diffeomorphisms, the utility of residual LGTs and CFC in this particular context is one of convenience, not necessity. Indeed, these transformations make some of the terms composing an observable vanish, in particular some long-wavelength effects within the finite volume relevant for observation, since they appear as the first few terms in an expansion in $x^i$. This also provides physical transparency, because in these coordinates one sees how short-wavelength processes decouple from metric long-wavelength fluctuations, i.e. an explicit manifestation of the equivalence principle. The price to pay is that the relation to Fourier space and the associated scalar-vector-tensor decomposition is complicated by the expansion in the spatial coordinates. 

In generic coordinates, such as those typically used in cosmology (e.g. conformal-Newtonian), many of the aforementioned long-wavelength coordinate artifacts are present, but are compensated by other terms in the full expression of the observable, {\it hence the importance of taking all terms into account for reliable results in general}. A particular example are the contributions to the observable from the observer point, which vanish in a CFC centered at the observer, but are needed to properly cancel out infrared-sensitive effects in generic gauge, as shown explicitly in \cite{Dai:2013kra,Biern:2016kys,Scaccabarozzi:2018vux,Grimm:2020ays,Baumgartner:2020llb,Castorina:2021xzs,Yoo:2022klz} and discussed in detail in \footnote{E. Mitsou and J. Yoo, The infrared (in)sensitivity of relativistic cosmological observable statistics, in preparation.} (for the statistical treatment of observer terms, see \cite{Mitsou:2019ocs}). Put differently, neglecting observer terms can lead to spurious infrared effects, which can therefore be misinterpreted as a relativistic signal. Given the flurry of upcoming surveys devoted to the observation of the very large scales, which will provide a window into primordial non-Gaussianity, a rigorous treatment of the infrared is now relevant.

{\bf A global SGT for a locally CFC metric} -- Finally, let us point out that the property of practical interest for using CFC, i.e. equation \eqref{eq:CFC}, can be achieved with a SGT. More precisely, starting with an arbitrary line-element with globally small fluctuations, we can build a (global) SGT that leads to equation \eqref{eq:CFC} and with corrections that kick in at the same distance, but differs from CFC at higher orders in $x^i$. One can therefore have the advantages of CFC within the coordinate subclass of cosmological perturbation theory, thus preserving SGT-invariant quantities in particular. 

Consider the coordinate transformation from a global system (e.g. conformal-Newtonian gauge) to the CFC associated with some world-line and express it as a deviation from the identity $x^{\mu} \to x^{\mu} + \xi^{\mu}(x)$. The metric coefficient functions in CFC coordinates are thus related to the original ones through
\bea
g_{\mu\nu}^{\rm CFC}(x) & = & \( \de_{\mu}^{\ro} + \frac{\pa \xi^{\ro}(x)}{\pa x^{\mu}} \) \( \de_{\nu}^{\si} + \frac{\pa \xi^{\si}(x)}{\pa x^{\nu}} \) \label{eq:gCFCtrans} \\
& & \times \sum_{n = 0}^{\infty} \frac{1}{n!} \[ \xi^{\mu_1} \dots \xi^{\mu_n} \frac{\pa}{\pa x^{\mu_1}} \dots \frac{\pa}{\pa x^{\mu_n}} \, g_{\ro\si} \](x) \, , \nn
\eea
where we have expanded the argument of the latter. The function $\xi^{\mu}$ is built order by order in the spatial distance to the world-line
\beq \label{eq:xiCFC}
\xi^{\mu}(x) = \sum_{n=0}^{\infty} \frac{1}{n!} \, f^{\mu}_{i_1 \dots i_n}(t) \, x^{i_1} \dots x^{i_n} \, ,
\eeq
and we note that the property \eqref{eq:CFC} involves only the terms up to $\Ord(\vec{x}^2)$, since \eqref{eq:gCFCtrans} contains at most first derivatives of $\xi^{\mu}$. Consequently, \eqref{eq:CFC} still holds if we use instead any $\xi^{\mu}(x)$ function with the same expansion up to $\Ord(\vec{x}^2)$. A simple example is given by
\bea 
\xi^{\mu}(x) & = & \[ f^{\mu}(t) + f_i^{\mu}(t) \, x^i + \frac{1}{2} \( f_{ij}^{\mu}(t) + \La_{ij}(t) \, f^{\mu}(t) \) x^i x^j \] \nn \\
& & \times \exp \[ - \frac{1}{2} \, \La_{ij}(t) \, x^i x^j \] \, , \label{eq:ffin}
\eea
where the $\La_{ij}(t)$ are fixed functions forming a matrix with positive-definite spectrum, e.g. $\La_{ij} = \de_{ij} / L^2$ for some length scale $L$. Indeed, \eqref{eq:ffin} corresponds to a smooth and globally-defined coordinate transformation leading to equation \eqref{eq:CFC}. However, if $\La_{ij}$ is too small, then $\pa_{\nu} \xi^{\mu}$ can become too large for some $x^i$ values for the corresponding gauge transformation to qualify as ``small''. We should therefore choose the length scales in $\La_{ij}$ to be small enough, if we want a practically useful transformation. These scales are given by ratios of the $f^{\mu}_{\dots}$ coefficients, or alternatively, the typical spatial variation of the original metric functions around the world-line. As a result, the $\Ord(\vec{x}^2)$ corrections of the resulting metric functions will kick in at the same typical distance as in the CFC case \eqref{eq:CFC}. Thus, for all practical purposes, the transformation given by \eqref{eq:ffin} with appropriate $\La_{ij}(t)$ functions has the same properties as CFC, but is a well-defined SGT. In particular, the function \eqref{eq:ffin} has a smooth Fourier transform, so the corresponding metric variation has a unique scalar-vector-tensor decomposition.

\begin{acknowledgments}
EM is supported by a Forschungskredit Grant of the University of Zurich (grant FK-21-130) and JY is supported by a Consolidator Grant of the European Research Council (ERC-2015-CoG grant 680886). 
\end{acknowledgments}

\clearpage

\bibliography{mybib}

\end{document}

%% file: mydefs.tex
\newcommand{\beq}{\begin{equation}}
\newcommand{\eeq}{\end{equation}}
\newcommand{\bea}{\begin{eqnarray}}
\newcommand{\eea}{\end{eqnarray}}
\newcommand{\ben}{\begin{enumerate}}
\newcommand{\een}{\end{enumerate}}

\newcommand{\pa}{\partial}

\newcommand{\Ord}{{\cal O}}

\renewcommand\({\left(}
\renewcommand\){\right)}
\renewcommand\[{\left[}
\renewcommand\]{\right]}
\newcommand{\bra}{\langle}
\newcommand{\ket}{\rangle}

\newcommand{\nn}{\nonumber}

\newcommand{\ga}{\gamma}

\newcommand{\de}{\delta}

\newcommand{\ze}{\zeta}
\newcommand{\et}{\eta}

\newcommand{\la}{\lambda}
\newcommand{\La}{\Lambda}
\newcommand{\ro}{\rho}
\newcommand{\si}{\sigma}

\newcommand{\ta}{\tau}

